\documentstyle[11pt]{article}

\begin{document}

\centerline{\Large Spectra of Free Diquark in the Bethe-Salpeter
Approach}

\vspace{1cm}

Yan-Ming Yu$^1$, Hong-Wei Ke$^1$, Yi-Bing Ding$^2$, Xin-Heng
Guo$^3$, Hong-Ying Jin$^4$, Xue-Qian Li$^1$, Peng-Nian Shen$^5$
and Guo-Li Wang$^6$

\vspace{0.5cm}

1. Department of Physics, Nankai University, Tianjin 300071,
China.

2. Department of Physics, The Graduate University of Chinese Academy
of Sciences, Beijing, 100049, China.

3. Institute of Low Energy Nuclear Physics, Beijing Normal
University, Beijing 100875, China.

4. The Institute of Modern Physics, Zhejiang University, Hangzhou,
310027, China.

5. Institute of High Energy Physics, Chinese Academy of Sciences,
P.O. Box 918-4, Beijing 100049, China.

6. Department of Physics, The Harbin Institute of Technology,
Harbin, 150001, China.

\vspace{1cm}

\begin{center}
\begin{minipage}{12cm}
\noindent Abstract:

In this work, we employ the Bethe-Salpeter (B-S) equation to
investigate the spectra of free diquarks and their B-S wave
functions. We find that the B-S approach can be consistently
applied to study the diqaurks with two heavy quarks or one heavy
and one light quarks, but for two light-quark systems, the results
are not reliable. There are a few free parameters in the whole
scenario which can only be fixed phenomenologically. Thus, to
determine them, one has to study baryons which are composed of
quarks and diquarks.

\end{minipage}
\end{center}

\vspace{1cm}

\section{Introduction}
The subject of diquarks has attracted attentions of theorists of
high energy physics for decades. The reason is obvious. Since
baryons are composed of three valence quarks, the three-body
system is much more complicated than mesons which are two-body
systems of quark and antiquark. If the diquark picture is applied,
namely two of the constituent quarks constitute a
color-anti-triplet sub-system, the three-body problem can be
reduced to a much simpler two-body system. However, one would ask
whether QCD, which is responsible for interaction between quarks
or antiquarks, favors such diquark structure of two constituent
quarks.

Recently, the topic on diquarks revives because it may bring up
some direct phenomenological consequences. At relativistic  heavy
ion collisions (RHIC) with high temperature and density, diquark
production and even diquark condensation become hot topics
\cite{RHIC}. Some authors suggest that pentaquarks are of a
diquark-diquark-anti-quark structure
\cite{jandw,jaffe,wilczek,shuryak}, and Zou et al. proposed that
there is a pentaquark component in nucleons \cite{Zou}. All the
subjects concern a dynamics which results in the substantial
diquark structure.

In 1964 Gell-Mann first proposed feasibility of the diquark
structure \cite{gellman}, then Ida, Kabayashi\cite{ida}, Tassie et
al.\cite{lichtenberg} applied the concept to study baryons. Later
many authors carefully analyzed the quantum numbers based on the
group theory \cite{bose2,miyazawa1,miyazawa2}.
Anselmino\cite{Anselmino} et al. indicated that two heavy quarks or
one heavy and one light quarks may constitute stable scalar or axial
vector diquark of color anti-triplet\cite{jaffe,wilczek}. However,
for the two light quarks, it is not very clear if a substantial
diquark can exist.

In this work we employ the Bethe-Salpeter (B-S) equation to study
the spectra of free diqaurks which are composed of two heavy
quarks or one-heavy-one-light quarks, we also try to extend this
approach to study the diquarks of two light quarks. Our results
show that just as for the light pseudoscalar pions, this framework
does not work well, unless one considers extra contributions to
quark masses. Dai et al. \cite{Dai} used the method to study the
pion structure and recently, Wang et al. \cite{Wang,Wang2}
employed the same approach to study the diquarks which contain
only two light quarks. We will come back to this issue in the last
section. While numerically solving the B-S equation, we have
employed the method developed by Chang et al. \cite{Chang}, which
is proved to be powerful and efficient in the numerical
computations.

This work is organized as follows. After this short introduction,
we derive our formulations. In Sec.III, we present the numerical
results, and then in the last section, we briefly discuss our
results.

\section{Formulation}

In this work, we use the B-S equations to study the diquark
structure. First we construct the Green's function of the diquark in
the $4\times 4$ matrix form and derive the corresponding B-S
equation.

\subsection{Structures of diquarks}

One can define the B-S wave function of a diquark as
\cite{origin,shuryak}
\begin{equation}\label{green1}\chi^k_P(x_1,x_2)=\frac{\epsilon_{ijk}}{\sqrt{6}}<0|T(\psi_i(x_1)\overline{\psi_j^c}(x_2))|D>=e^{-iP\cdot
X}\chi^k_P(x),\end{equation} where $i,\; j,\; k$ are color indices
which will be omitted later without misunderstanding. A Fourier
transformation brings it into the momentum space as
\begin{equation}\chi^k_P(x)=\frac{1}{(2\pi)^4}\int d^4q e^{iq\cdot
x}\chi^k_P(p),
\end{equation}
where
$$\psi^c(x)=C\bar{\psi}^T ,C=i\gamma_0\gamma_2 ,X=\alpha_1x_1+\alpha_2x_2,x=x_1-x_2,
$$\begin{equation}\alpha_1=\frac{m_1}{m_1+m_2} ,\alpha_2=\frac{m_2}{m_1+m_2} ,P=p_1+p_2
,q=\alpha_2p_1-\alpha_1p_2.
\end{equation}
Thus the B-S wavefunction in the momentum space reads
\begin{equation}\label{diqbs}
(p\!\!\!\slash_1-m_1)\chi_P(q)(p\!\!\!\slash_2+m_2)=\int
d^4k\overline{V}(P,q;k)\chi_P(k),
\end{equation}
where $\overline{V}(P,q;k)$ is the kernel. Although the B-S
equation has the same form as that for mesons, the kernels for the
two cases are different. For the diquarks there are no
annihilation diagrams which exist for meson case. In fact, the
definition of the diquark B-S wavefuntion automatically eliminates
annihilation diagrams\footnote{ The four-point Green's function
for diquark is defined  as
$<0|T(\psi(x_1)\bar\psi^c(x_2)\psi^c(x_3)\bar\psi(x_4)|0>$ whreas
for meson it is
$<0|T(\psi(x_1)\bar\psi(x_2)\psi(x_3)\bar\psi(x_4)|0>$, since
$\psi(x)$ and $\bar\psi^c(y)$ cannot contract, the annihilation
diagrams do not exist in the diquar case. }. Greiner
\cite{greiner} indicates that the B-S equations for quark-quark
system (diquark) and for quark-antiquark system (meson) have a
formal symmetry.

According to the method given in Ref. \cite{zhang1}, one can
obtain an equation group which contains four independent
equations.

\begin{eqnarray}
\label{eq1}
(M-\omega_{1P}-\omega_{2P})\varphi^{++}_{P}(q^\mu_{P_\perp})=
\Lambda^{+}_{1P}(q^\mu_{P_\perp})
\eta_{P}(q^\mu_{P_\perp})\Lambda^{+}_{2P}(q^\mu_{P_\perp})\,,
\end{eqnarray}
\begin{eqnarray}
\label{eq2}
(M+\omega_{1P}+\omega_{2P})\varphi^{--}_{P}(q^\mu_{P_\perp})=-
\Lambda^{-}_{1P}(q^\mu_{P_\perp})
\eta_{P}(q^\mu_{P_\perp})\Lambda^{-}_{2P}(q^\mu_{P_\perp})\,,
\end{eqnarray}
and
\begin{equation}
\label{condition1} \varphi^{+-}_{P}(q^\mu_{P_\perp})=0,
\end{equation}
\begin{equation}
\label{condition2} \varphi^{-+}_{P}(q^\mu_{P_\perp})=0.
\end{equation}
The  B-S wavefunctions are normalized as  following:
\begin{equation}\int\frac{d^3q_{P\perp}}{(2\pi)^3}[\overline{\varphi}^{++}(q_{P\perp})
\frac{P\!\!\!\slash}{M}\varphi^{++}(q_{P\perp})\frac{P\!\!\!\slash}{M}-\overline{\varphi}^{--}(q_{P\perp})
\frac{P\!\!\!\slash}{M}\varphi^{--}(q_{P\perp})\frac{P\!\!\!\slash}{M}]=2M.
\end{equation}

The kernel for the B-S equation for diquark has been widely
discussed by many authors\cite{wang1}
\begin{equation}I(r)=V_s(r)+V_0+\gamma^\mu\otimes\gamma_\mu V_v(r)=\beta\lambda
r+V_0-\gamma^\mu\otimes\gamma_\mu \frac{2}{3}\frac{\alpha_s}
{r}.\end{equation} To avoid the infrared divergence, one can
introduce a convergence factor $e^{-\alpha r}$. Thus the
potentials are respectively\footnote{In fact, for the Coulomb
term, there is no infrared divergence, but the linear confinement,
which is in the form of $1/{\bf k}^4$ in the momentum space,
diverges. Introduction of a converging factor is necessary and is
benign for the Coulomb term.}
\begin{equation}V_s(r)=\frac{\beta\lambda}{\alpha}(1-e^{-\alpha
r}),\end{equation}and
\begin{equation}V_v(r)=-\frac{2}{3}\frac{\alpha_s}{r}e^{-\alpha r}. \end{equation}
In the momentume space the potential reads
\begin{equation}K({\bf q})=I\otimes
IV_s({\bf q})+\gamma^\mu\otimes\gamma_\mu V_v({\bf q})\end{equation}
where
\begin{equation}V_s({\bf q})=-(\frac{\beta\lambda}{\alpha}+V_0)\delta^3({\bf q})+
\frac{\beta\lambda}{\pi^2}\frac{1}{({\bf
q}^2+\alpha^2)^2},\end{equation} and
\begin{equation}V_v({\bf q})=-\frac{1}{3\pi^2}\frac{\alpha_s({\bf q})}{({\bf q}^2+\alpha^2)}.\end{equation}
The running coupling constant $\alpha_s({\bf q})$ is
\begin{equation}\alpha_s({\bf q})=\frac{4\pi}{9}\frac{1}{ln(a+\frac{{\bf q}^2}{\Lambda^2_{QCQ}})},\end{equation}
where $a$ is a constant which freezes the running coupling
constant at low momentum.

Because diquark exists in the color-$\bar 3$ state, $<\bar
3|\lambda^a\lambda^a|\bar 3>={1\over 2}<0|\lambda^a\lambda^a|0>$,
where $|0>$ is the color singlet. Thus the coefficient of the
Coulomb term, which originates from the one-gluon exchange, is
$-\frac{2\alpha_s}{3}$ for diquarks, whereas it is
$-\frac{4\alpha_s}{3}$ for mesons. The linear potential comes from
the non-perturbative effects of QCD, it might have different form
for diquark and meson, and so far we cannot determine its exact
value from the QCD theory. Thus we introduce a phenomenological
parameter $\beta$. Indeed, some authors\cite{Wang} argued that for
diquarks, the linear confinement might not exist at all, if so,
$\beta=0$. On other aspect, if the dynamics for the bound state of
diquark is similar to that for mesons, one should expect
$\beta=0.5$, in analog to the coefficient ratio in front of the
Coulomb term. We set the parameter to be free   within a range of
$0\leq\beta\leq 1.0$.

\subsection{Scalar diquark system}

We carry out all the computations in the center-of-mass frame of
the diquark, thus $P=(M, {\bf 0})$ where $P$ is the total
four-momentum of the diquark and M is its mass.

For $0^{+}$ diquark, the general form of its wavefunction is
\begin{equation}\varphi({\bf q})=\gamma_0b_1({\bf q})+b_2({\bf q})+q\!\!\!\slash_{\perp}b_3({\bf q})
+\gamma_0q\!\!\!\slash_{\perp}b_4({\bf q}).
\end{equation}
Here we define $q_{\perp}$ as $(0,{\bf q})$ which is perpendicular
to $P$. Substituting the wavefunction into the equations
(\ref{condition1}) and (\ref{condition2}), we can obtain the
constraint conditions
\begin{equation}b_1({\bf q})=\frac{{\bf q}^2(m_1+m_2)^2b_4({\bf q})}{(\omega_1+\omega_2)({\bf q}^2-m_1m_2-\omega_1\omega_2)}
,\end{equation} and
\begin{equation}
b_2({\bf q})=\frac{{\bf q}^2(m_1+m_2)b_3({\bf q})}{{\bf
q}^2-m_1m_2-\omega_1\omega_2}.\end{equation}

With these constraints, the B-S wavefunction of a $0^+$ diquark
has the following form
\begin{equation}\varphi({\bf q})=\frac{{\bf q}^2(\omega_1-\omega_2)^2\gamma_0b_4({\bf q})}{(m_1-m_2)
({\bf q}^2-m_1m_2-\omega_1\omega_2)}+\frac{{\bf
q}^2(m_1+m_2)b_3({\bf q})}{{\bf q}^2-m_1m_2-\omega_1\omega_2}
+q\!\!\!\slash_{\perp}b_3({\bf q})+\gamma_0
q\!\!\!\slash_{\perp}b_4({\bf q}).\end{equation}

Substituting them into eqs.(\ref{eq1}) and (\ref{eq2}), we obtain
an equation group for the component functions:
$$(M-\omega_1-\omega_2)\frac{2{\bf q}^2[({\bf q}^2-m_1m_2-\omega_1\omega_2)b_4({\bf q})-(m_2\omega_1+m_1\omega_2)b_3({\bf q})]
}{{\bf q}^2-m_1m_2-\omega_1\omega_2}$$
$$=-\int\frac{d^3k}{(\omega_1\omega_2)({\bf k}^2-m_1m_2-\omega_{1k}\omega_{2k})}\left\{b_3({\bf k})[(V_s+4V_v)(m_1+m_2)
(\omega_1+\omega_2){\bf q}^2{\bf k}^2\right.$$
$$\left.+(V_s-2V_v)(m_1\omega_2+m_2\omega_1)({\bf k}^2-m_1m_2-\omega_{1k}\omega_{2k}){\bf q}\cdot{\bf k}]\right.$$
\begin{equation}\left.+b_4({\bf k})[(V_s-2V_v)(\omega_{1k}-\omega_{2k})^2{\bf q}^2{\bf k}^2
+V_s({\bf q}^2+m_1m_2+\omega_1\omega_2)({\bf
k}^2-m_1m_2-\omega_{1k}\omega_{2k}){\bf q}\cdot{\bf
k}]\right\},\end{equation}

$$\left(M+\omega_1+\omega_2\right)\frac{2{\bf q}^2[({\bf q}^2-m_1m_2-\omega_1\omega_2)b_4({\bf q})+(m_2\omega_1+m_1\omega_2)b_3({\bf q})]
}{{\bf q}^2-m_1m_2-\omega_1\omega_2}$$
$$=\int\frac{d^3k}{(\omega_1\omega_2)({\bf k}^2-m_1m_2-\omega_{1k}\omega_{2k})}\left\{-b_3({\bf k})[(V_s+4V_v)(m_1+m_2)
(\omega_1+\omega_2){\bf q}^2\vec{k^2}\right.$$
$$\left.+(V_s-2V_v)(m_1\omega_2+m_2\omega_1)({\bf k}^2-m_1m_2-\omega_{1k}\omega_{2k}){\bf q}\cdot{\bf k}]\right.$$
\begin{equation}\left.+b_4({\bf k})[(V_s-2V_v)(\omega_{1k}-\omega_{2k})^2{\bf q}^2{\bf k}^2
+V_s({\bf q}^2+m_1m_2+\omega_1\omega_2)({\bf
k}^2-m_1m_2-\omega_{1k}\omega_{2k}){\bf q}\cdot{\bf
k}]\right\}.\end{equation}

The normalization of the component functions is set as
\begin{equation}
\int\frac{d^3q}{(2\pi)^3}\frac{16\omega_1\omega_2(m_1+m_2)b_3({\bf
q})b_4({\bf q})} {(\omega_1+\omega_2)(-{\bf
q}^2+m_1m_2+\omega_1\omega_2)}=2M.\end{equation}

\subsection{Axial Vector diquark system}
The general form of the wavefunction of $1^+$ diquark reads as
$$\varphi({\bf q})=\left\{q_\perp\cdot\epsilon^\lambda_\perp[f_1({\bf q})+\gamma_0f_2({\bf q})+\frac{q\!\!\!\slash_\perp
 f_3({\bf q})}{M}+\frac{\gamma_0q\!\!\!\slash_\perp
 f_4({\bf q})}{M}]\right.$$
\begin{equation}\left.+M\epsilon\!\!\!\slash^\lambda
f_5({\bf q})+M\epsilon\!\!\!\slash^\lambda\gamma_0f_6({\bf
q})+q\!\!\!\slash\epsilon\!\!\!\slash^\lambda f_7({\bf
q})+i\epsilon^{0ijk}\epsilon^\lambda_iq_{\perp
j}\gamma_k\gamma_5f_8({\bf q})]\right\}\gamma_5.
\end{equation}
The constraint conditions are
$$f_2({\bf q})=\frac{\omega_2-\omega_1}{\omega_1+\omega_2}f_8({\bf q})+\frac{{\bf q}^2-m_1m_2-\omega_1
\omega_2}{M(m_1+m_2)}f_4({\bf q}),\;\;\;\;\; f_6({\bf
q})=\frac{m_1\omega_2-m_2\omega_1}{M(\omega_1+\omega_2)}f_8({\bf
q}),$$
\begin{equation}f_3({\bf q})=\frac{M[2M\omega_1f_5({\bf q})+(m_2\omega_1-m_1\omega_2)f_1({\bf q})]}
{{\bf q}^2(\omega_1+\omega_2)},\;\;\;\; f_7({\bf
q})=-\frac{M(m_1\omega_2 +m_2\omega_1)}{{\bf
q}^2(\omega_1+\omega_2)}f_5({\bf q}).\end{equation}

With these conditions the B-S wavefunction of $1^+$ diquark can be
further written as
$$\varphi({\bf q})=\{q_\perp\cdot\epsilon^\lambda_\perp[f_1({\bf q})+\gamma_0\frac{\omega_2-\omega_1}
{\omega_1+\omega_2}f_8({\bf q})+\gamma_0\frac{{\bf
q}^2-m_1m_2-\omega_1 \omega_2}{M(m_1+m_2)}f_4({\bf
q})$$$$+\frac{q\!\!\!\slash_\perp}{M}\frac{M[2M\omega_1f_5({\bf
q})+(m_2\omega_1-m_1\omega_2)f_1({\bf q})]} {{\bf
q}^2(\omega_1+\omega_2)}+\frac{\gamma_0q\!\!\!\slash_\perp f_4({\bf
q})}{M}]$$$$+M\epsilon\!\!\!\slash^\lambda_\perp f_5({\bf
q})+M\epsilon\!\!\!\slash^\lambda_\perp\gamma_0 \frac{{\bf
q}^2+m_1m_2-\omega_1\omega_2}{M(m_2-m_1)}f_8({\bf q})$$
\begin{equation}-q\!\!\!\slash_\perp\epsilon\!\!\!\slash^\lambda_\perp\frac{M(m_1\omega_2
+m_2\omega_1)}{{\bf q}^2(\omega_1+\omega_2)}f_5({\bf
q})+i\epsilon^{0ijk}\epsilon^\lambda_iq_{\perp
j}\gamma_k\gamma_5f_8({\bf q})]\}\gamma_5.\end{equation} The
coupled equations are
$$2(M-\omega_1-\omega_2)[\frac{M{\bf q}^2(\omega_1+\omega_2)f_1({\bf q})-M^2(m_1\omega_2+m_2\omega_1)f_5({\bf q})}
{\omega_1+\omega_2}$$$$-\frac{({\bf
q}^2+m_1m_2+\omega_1\omega_2){\bf q}^2f_4({\bf q})+M(m_1-m_2){\bf
q}^2f_8({\bf q})} {\omega_1+\omega_2}]$$$$=\int\frac{d^3q}{{\bf
k}^2\omega_1\omega_2(\omega_{1k}+\omega_{2k})}\{f_1({\bf
q})M[(m_1-m_2) (V_s+2V_v)(m_1\omega_{2k}-m_2\omega_{1k})({\bf
q}\cdot{\bf k})^2$$$$-(V_s-4V_v)({\bf q}^2+m_1m_2+\omega_1\omega_2)
(\omega_{1k}+\omega_{2k}){\bf k}^2{\bf q}\cdot{\bf k}]$$$$+f_4({\bf
q}){\bf k}^2[(V_s+2V_v)(m_1\omega_2+m_2\omega_1)
(m_1\omega_{2k}+m_2\omega_{1k}){\bf q}\cdot{\bf
k}-V_s(\omega_1+\omega_2)(\omega_{1k}+\omega_{2k})({\bf q}\cdot{\bf
k})^2 ]$$$$+f_5({\bf q})M^2[(m_1-m_2)(V_s+2V_v)({\bf q}^2{\bf
k}^2(\omega_{1k}+\omega_{2k})-2\omega_{1k}({\bf q}\cdot{\bf k})^2)
$$$$+(V_s-4V_v)({\bf q}^2+m_1m_2+\omega_1\omega_2)(m_1\omega_{2k}+m_2\omega_{1k}){\bf q}\cdot{\bf k}]$$\begin{equation}+f_8({\bf q})
M[V_s(\omega_1+\omega_2)(m_2\omega_{1k}-m_1\omega_{2k}){\bf
q}^2{\bf k}^2+(V_s+2V_v)(m_1\omega_2+m_2\omega_1)(\omega_{1k}
-\omega_{2k}){\bf k}^2{\bf q}\cdot{\bf k}]\},\end{equation}

$$2(M+\omega_1+\omega_2)[\frac{M{\bf q}^2(\omega_1+\omega_2)f_1({\bf q})-M^2(m_1\omega_2+m_2\omega_1)f_5({\bf q})}
{\omega_1+\omega_2}$$$$+\frac{({\bf
q}^2+m_1m_2+\omega_1\omega_2){\bf q}^2f_4({\bf q})+M(m_1-m_2){\bf
q}^2f_8({\bf q})} {\omega_1+\omega_2}]$$$$=\int\frac{d^3q}{{\bf
k}^2\omega_1\omega_2(\omega_{1k}+\omega_{2k})}\{-f_1({\bf
q})M[(m_1-m_2) (V_s+2V_v)(m_1\omega_{2k}-m_2\omega_{1k})({\bf
q}\cdot{\bf k})^2$$$$-(V_s-4V_v)({\bf q}^2+m_1m_2+\omega_1\omega_2)
(\omega_{1k}+\omega_{2k}){\bf k}^2{\bf q}\cdot{\bf k}]$$$$+f_4({\bf
q}){\bf k}^2[(V_s+2V_v)(m_1\omega_2+m_2\omega_1)
(m_1\omega_{2k}+m_2\omega_{1k}){\bf q}\cdot{\bf
k}-V_s(\omega_1+\omega_2)(\omega_{1k}+\omega_{2k})({\bf q}\cdot{\bf
k})^2 ]$$$$-f_5({\bf q})M^2[(m_1-m_2)(V_s+2V_v)({\bf q}^2{\bf
k}^2(\omega_{1k}+\omega_{2k})-2\omega_{1k}({\bf q}\cdot{\bf k})^2)
$$$$+(V_s-4V_v)({\bf q}^2+m_1m_2+\omega_1\omega_2)(m_1\omega_{2k}+m_2\omega_{1k}){\bf q}\cdot{\bf k}]$$\begin{equation}+f_8({\bf q})
M[V_s(\omega_1+\omega_2)(m_2\omega_{1k}-m_1\omega_{2k}){\bf
q}^2{\bf k}^2+(V_s+2V_v)(m_1\omega_2+m_2\omega_1)(\omega_{1k}
-\omega_{2k}){\bf k}^2{\bf q}\cdot{\bf k}]\},\end{equation}

$$4(M-\omega_1-\omega_2)\frac{M^2({\bf q}^2+m_1m_2+\omega_1\omega_2)f_5({\bf q})-M(\omega_1+\omega_2){\bf q}^2f_8({\bf q})}
{\omega_1+\omega_2}$$$$=\int\frac{d^3q}{{\bf
k}^2\omega_1\omega_2(\omega_{1k}+\omega_{2k})}\{-f_1({\bf
q})M(V_s+2V_v)
(\omega_1+\omega_2)(m_2\omega_{1k}-m_1\omega_{2k})[{\bf q}^2{\bf
k}^2-({\bf q}\cdot{\bf k})^2]$$$$+f_4({\bf q}){\bf k}^2(
m_1-m_2)V_s(\omega_{1k}+\omega_{2k})[({\bf q}\cdot{\bf k})^2-{\bf
q}^2{\bf k}^2]$$$$+2M^2f_5({\bf q})[(V_s+2V_v)
(\omega_1+\omega_2)(\omega_{2k}{\bf q}^2{\bf k}^2+\omega_{1k}({\bf
q}\cdot{\bf k})^2)-V_s(m_1\omega_2+m_2\omega_1)
(m_1\omega_{1k}+m_2\omega_{2k}){\bf q}\cdot{\bf
k}]$$\begin{equation}+2M{\bf k}^2f_8({\bf q})[(V_s-2V_v)({\bf
q}^2+m_1m_2 +\omega_1\omega_2)(\omega_{1k}+\omega_{2k}){\bf
q}\cdot{\bf k}-V_s(m_1-m_2)(m_1\omega_{2k}-m_2\omega_{1k}){\bf
q}^2]\end{equation}

$$4(M+\omega_1+\omega_2)\frac{-M^2({\bf q}^2+m_1m_2+\omega_1\omega_2)f_5({\bf q})-M(\omega_1+\omega_2){\bf q}^2f_8({\bf q})}
{\omega_1+\omega_2}$$$$=\int\frac{d^3q}{{\bf
k}^2\omega_1\omega_2(\omega_{1k}+\omega_{2k})}\{-f_1({\bf
q})M(V_s+2V_v)
(\omega_1+\omega_2)(m_2\omega_{1k}-m_1\omega_{2k})[{\bf q}^2{\bf
k}^2-({\bf q}\cdot{\bf k})^2]$$$$-f_4({\bf q}){\bf k}^2(
m_1-m_2)V_s(\omega_{1k}+\omega_{2k})[({\bf q}\cdot{\bf k})^2-{\bf
q}^2{\bf k}^2]$$$$+2M^2f_5({\bf q})[(V_s+2V_v)
(\omega_1+\omega_2)(\omega_{2k}{\bf q}^2{\bf k}^2+\omega_{1k}({\bf
q}\cdot{\bf k})^2)-V_s(m_1\omega_2+m_2\omega_1)
(m_1\omega_{1k}+m_2\omega_{2k}){\bf q}\cdot{\bf
k}]$$\begin{equation}-2M{\bf k}^2f_8({\bf q})[(V_s-2V_v)({\bf
q}^2+m_1m_2 +\omega_1\omega_2)(\omega_{1k}+\omega_{2k}){\bf
q}\cdot{\bf k}-V_s(m_1-m_2)(m_1\omega_{2k}-m_2\omega_{1k}){\bf
q}^2].\end{equation}

The normalization is set as
$$\int\frac{d^3q}{(2\pi)^3M(\omega_1+\omega_2)^2}16[\omega_1\omega_2{\bf q}^2(\omega_1+\omega_2)f_1({\bf q})f_4({\bf q})
$$$$-\omega_1\omega_2(m_1\omega_2+m_2\omega_1)Mf_4({\bf q})f_5({\bf q})+\omega_1\omega_2(m_1\omega_2-m_2\omega_1)f_1({\bf q})
f_8({\bf
q})$$$$+M^2(m_2^2\omega_1-m_1m_2\omega_1-m_1m_2\omega_2+\omega_1\omega_2^2-\omega_2{\bf
q}^2)f_5({\bf q})f_8({\bf q})] =2M.$$

\section{Numerical results and discussions}
In our numerical computations, the input parameters\cite{wang1} are
 $m_u=0.305\; {\rm GeV},\;\;  m_d=0.311\; {\rm GeV},\;\; m_s=0.487 \;
{\rm GeV},\;\;  m_c=1.7553\; {\rm GeV},\;\; m_b=5.224\; {\rm
GeV},\;\; \lambda=0.20\; {\rm GeV}^2 , \Lambda_{QCD}=0.26\; {\rm
GeV},\;\; a=2.71828\;\;,{\rm and }\;\; \alpha=0.06\; {\rm GeV}$.

Besides the spectra of different free diquark states, we also
evaluate their mean square-root radius which is defined as
\begin{equation}\sqrt{<r^2>}=\sqrt{<r^2_x>+<r^2_y>+<r^2_z>},
\end{equation}
with
\begin{equation}
<r^2_x>=\frac{\int d^3qTr[<\varphi({\bf
q})|-\frac{\partial^2}{\partial q^2_x}|\varphi({\bf q})>]} {\int
d^3q Tr[<\varphi({\bf q})|\varphi({\bf q})>]}.
\end{equation}

Our numerical results are tabulated below. In the following
tables, we present the results corresponding to different values
of $\beta$ and the vacuum expectation value $V_0$ (or the
zero-point value).

\begin{center}
\begin{tabular}{|c|c|c|c|c|c|c|}\hline
\multicolumn{1}{|c|}{$V_0(GeV)$}&\multicolumn{2}{|c|}{0.0}&\multicolumn{2}{|c|}{-0.3}&\multicolumn{2}{|c|}{-0.6}\\\hline
&mass&msr&mass&msr&mass&msr
\\\hline$(ud)_{0^+}$&0.6032&3.5287&0.3216&11.3828&-&-
\\\hline$(us)_{0^+}$&0.7839&3.6089&0.4962&11.3616&0.1982&11.3802
\\\hline$(ds)_{0^+}$&0.7898&3.6330&0.5021&11.3597&0.2041&11.3790
\\\hline$(uc)_{0^+}$&2.057&5.2610&1.763&11.3336&1.465&11.3652
\\\hline$(dc)_{0^+}$&2.063&5.5278&1.769&11.3293&1.471&11.3639
\\\hline$(sc)_{0^+}$&2.230&2.8423&1.943&10.0198&1.644&11.2610
\\\hline$(ub)_{0^+}$&5.527&5.2038&5.232&11.3260&4.933&11.3615
\\\hline$(db)_{0^+}$&5.532&6.4324&5.238&11.3217&4.939&11.3594
\\\hline$(bs)_{0^+}$&5.698&2.7435&5.410&8.8897&5.114&10.8857
\\\hline$(bc)_{0^+}$&6.918&1.1725&6.627&1.3093&6.334&1.4472
\\\hline$(uu)_{1^+}$&0.6127&11.2596&0.3156&11.3421&-&-
\\\hline$(ud)_{1^+}$&0.6186&11.2483&0.3215&11.3416&-&-
\\\hline$(dd)_{1^+}$&0.6246&11.2458&0.3275&11.3411&-&-
\\\hline$(us)_{1^+}$&0.7940&11.2386&0.4962&11.2577&0.1980&11.0937
\\\hline$(ds)_{1^+}$&0.8000&11.0985&0.5021&10.5140&0.2038&11.2139
\\\hline$(ss)_{1^+}$&0.9725&5.5968&0.6767&11.2590&0.3778&11.2756
\\\hline$(uc)_{1^+}$&2.060&5.4201&1.763&4.0055&1.465&3.8727
\\\hline$(dc)_{1^+}$&2.066&5.2305&1.769&3.9799&1.471&3.8273
\\\hline$(sc)_{1^+}$&2.232&2.9041&1.942&7.9565&1.644&3.5679
\\\hline$(cc)_{1^+}$&3.460&1.3865&3.169&1.5507&2.876&1.7133
\\\hline$(ub)_{1^+}$&5.528&4.3950&5.232&11.2767&4.933&4.8772
\\\hline$(db)_{1^+}$&5.534&4.2068&5.238&5.0016&4.939&4.7222
\\\hline$(bs)_{1^+}$&5.700&2.4584&5.411&4.9070&5.113&11.1654
\\\hline$(bc)_{1^+}$&6.912&1.1162&6.621&1.2287&6.328&1.3427
\\\hline$(bb)_{1^+}$&10.33&0.7254&10.04&0.7500&9.740&0.7742
\\\hline
\end{tabular}
\begin{table}[h]
  \centering
  \caption{diquark mass(GeV) and mean square-root radius (fm) with $\beta=0.0$}\label{tab7}
\end{table}
\end{center}
\vspace{0.3cm}

\begin{center}
\begin{tabular}{|c|c|c|c|c|c|c|}\hline
\multicolumn{1}{|c|}{$V_0(GeV)$}&\multicolumn{2}{|c|}{0.0}&\multicolumn{2}{|c|}{-0.3}
&\multicolumn{2}{|c|}{-0.6}\\\hline &mass&msr&mass&msr&mass&msr
\\\hline$(ud)_{0^+}$&0.7397&4.0056&0.6724&2.0227&0.5139&2.3418
\\\hline$(us)_{0^+}$&0.9496&1.7019&0.8372&1.7572&0.6583&2.1020
\\\hline$(ds)_{0^+}$&0.9565&1.6778&0.8423&1.7470&0.6622&2.0898
\\\hline$(uc)_{0^+}$&2.274&1.1711&2.096&1.4230&1.889&1.7430
\\\hline$(dc)_{0^+}$&2.280&1.1652&2.100&1.4152&1.898&1.7300
\\\hline$(sc)_{0^+}$&2.445&1.0478&2.232&1.2454&1.999&1.4569
\\\hline$(ub)_{0^+}$&5.759&1.1265&5.569&1.3731&5.356&1.6822
\\\hline$(db)_{0^+}$&5.764&1.1206&5.572&1.3648&5.359&1.6686
\\\hline$(bs)_{0^+}$&5.920&0.9992&5.696&1.1815&5.458&1.3777
\\\hline$(bc)_{0^+}$&7.088&0.7089&6.810&0.7576&6.532&0.8039
\\\hline$(uu)_{1^+}$&0.8846&1.5497&0.6868&1.6647&0.4097&1.4848
\\\hline$(ud)_{1^+}$&0.8904&1.5471&0.6918&1.6598&0.4152&1.4856
\\\hline$(dd)_{1^+}$&0.8962&1.5376&0.6968&1.6538&0.4206&1.4857
\\\hline$(us)_{1^+}$&1.058&1.4241&0.8476&1.5606&0.5896&1.5354
\\\hline$(ds)_{1^+}$&1.064&1.4190&0.8521&1.5538&0.5935&1.5294
\\\hline$(ss)_{1^+}$&1.229&1.2974&0.9982&1.4166&0.7323&1.4287
\\\hline$(uc)_{1^+}$&2.299&1.2302&2.097&1.4450&1.877&1.7059
\\\hline$(dc)_{1^+}$&2.305&1.2252&2.101&1.4375&1.880&1.6925
\\\hline$(sc)_{1^+}$&2.462&1.0995&2.228&1.2476&1.982&1.4001
\\\hline$(cc)_{1^+}$&3.648&0.8163&3.371&0.8633&3.090&0.9053
\\\hline$(ub)_{1^+}$&5.764&1.1815&5.567&1.3683&5.350&1.6523
\\\hline$(db)_{1^+}$&5.769&1.1826&5.570&1.3887&5.353&1.6462
\\\hline$(bs)_{1^+}$&5.923&1.0507&5.694&1.2232&5.453&1.4060
\\\hline$(bc)_{1^+}$&7.078&0.7069&6.798&0.7468&6.517&0.7855
\\\hline$(bb)_{1^+}$&10.46&0.5329&10.17&0.5455&9.874&0.5579
\\\hline
\end{tabular}
\begin{table}[h]
  \centering
  \caption{diquark mass(GeV) and mean square root radius (fm) with $\beta=0.25$}\label{tab7}
\end{table}
\end{center}
\vspace{0.3cm}
\begin{center}
\begin{tabular}{|c|c|c|c|c|c|c|}\hline
\multicolumn{1}{|c|}{$V_0(GeV)$}&\multicolumn{2}{|c|}{0.0}&\multicolumn{2}{|c|}{-0.3}
&\multicolumn{2}{|c|}{-0.6}\\\hline &mass&msr&mass&msr&mass&msr
\\\hline$(ud)_{0^+}$&0.7904&1.3035&0.7911&2.0982&0.7105&1.7727
\\\hline$(us)_{0^+}$&1.028&2.3651&0.9731&1.5194&0.8582&1.5598
\\\hline$(ds)_{0^+}$&1.036&2.2344&0.9789&1.5046&0.8623&1.5515
\\\hline$(uc)_{0^+}$&2.394&0.9782&2.251&1.0897&2.086&1.2343
\\\hline$(dc)_{0^+}$&2.400&0.9725&2.255&1.0838&2.088&1.2277
\\\hline$(sc)_{0^+}$&2.575&0.8603&2.393&0.9666&2.192&1.0854
\\\hline$(ub)_{0^+}$&5.888&0.9047&5.728&1.0192&5.553&1.1626
\\\hline$(db)_{0^+}$&5.894&0.8995&5.733&1.0130&5.556&1.1560
\\\hline$(bs)_{0^+}$&6.056&0.7977&5.859&0.8979&5.648&1.0088
\\\hline$(bc)_{0^+}$&7.215&0.5893&6.946&0.6250&6.675&0.6574
\\\hline$(uu)_{1^+}$&0.9900&0.4521&0.8590&1.3209&0.6549&1.2716
\\\hline$(ud)_{1^+}$&0.9973&0.4806&0.8644&1.3176&0.6596&1.2697
\\\hline$(dd)_{1^+}$&1.004&0.5075&0.8697&1.3115&0.6642&1.2664
\\\hline$(us)_{1^+}$&1.192&0.7153&1.0222&1.2223&0.8099&1.2278
\\\hline$(ds)_{1^+}$&1.199&0.7244&1.027&1.2185&0.8139&1.2239
\\\hline$(ss)_{1^+}$&1.375&1.0575&1.178&1.1206&0.9470&1.1405
\\\hline$(uc)_{1^+}$&2.428&0.9021&2.252&1.0265&2.064&1.1565
\\\hline$(dc)_{1^+}$&2.434&0.9030&2.257&1.0251&2.067&1.1527
\\\hline$(sc)_{1^+}$&2.603&0.8764&2.391&0.9605&2.168&1.0456
\\\hline$(cc)_{1^+}$&3.789&0.6819&3.520&0.7152&3.247&0.7451
\\\hline$(ub)_{1^+}$&5.884&0.7958&5.717&0.9110&5.535&1.0536
\\\hline$(db)_{1^+}$&5.890&0.7996&5.721&0.9186&5.539&1.0403
\\\hline$(bs)_{1^+}$&6.056&0.8078&5.852&0.8995&5.636&1.0172
\\\hline$(bc)_{1^+}$&7.206&0.5962&6.933&0.6259&6.658&0.6527
\\\hline$(bb)_{1^+}$&10.56&0.4669&10.28&0.4765&9.985&0.4854
\\\hline
\end{tabular}
\begin{table}[h]
  \centering
  \caption{diquark mass(GeV) and mean square root radius (fm) with $\beta=0.5$}\label{tab7}
\end{table}
\end{center}
\vspace{0.3cm}
\begin{center}
\begin{tabular}{|c|c|c|c|c|c|c|}\hline
\multicolumn{1}{|c|}{$V_0(GeV)$}&\multicolumn{2}{|c|}{0.0}&\multicolumn{2}{|c|}{-0.3}
&\multicolumn{2}{|c|}{-0.6}\\\hline &mass&msr&mass&msr&mass&msr
\\\hline$(ud)_{0^+}$&-&-&0.4383&9.1040&0.6124&1.7486
\\\hline$(us)_{0^+}$&1.069&9.1714&1.059&1.6103&0.9856&1.4060
\\\hline$(ds)_{0^+}$&1.083&3.2217&1.065&1.5805&0.9904&1.3955
\\\hline$(uc)_{0^+}$&2.485&0.8955&2.364&0.9580&2.223&1.0424
\\\hline$(dc)_{0^+}$&2.492&0.8894&2.368&0.9524&2.226&1.0372
\\\hline$(sc)_{0^+}$&2.677&0.7734&2.516&0.8427&2.337&0.9227
\\\hline$(ub)_{0^+}$&5.987&0.8029&5.846&0.8744&5.692&0.9620
\\\hline$(db)_{0^+}$&5.993&0.7978&5.850&0.8692&5.695&0.9565
\\\hline$(bs)_{0^+}$&6.163&0.7012&5.984&0.7688&5.792&0.8451
\\\hline$(bc)_{0^+}$&7.324&0.5236&7.063&0.4804&6.797&0.5805
\\\hline$(uu)_{1^+}$&-&-&0.3729&10.0319&0.8123&1.1460
\\\hline$(ud)_{1^+}$&-&-&0.4387&9.9951&0.8172&1.1435
\\\hline$(dd)_{1^+}$&-&-&0.4961&9.9615&0.8220&1.1391
\\\hline$(us)_{1^+}$&1.069&10.0515&1.143&1.0528&0.9662&1.0805
\\\hline$(ds)_{1^+}$&1.099&10.0266&1.149&1.0528&0.9707&1.0776
\\\hline$(ss)_{1^+}$&1.481&0.9391&1.311&0.9839&1.108&1.0013
\\\hline$(uc)_{1^+}$&2.525&0.7164&2.366&0.8233&2.195&0.9183
\\\hline$(dc)_{1^+}$&2.532&0.7193&2.371&0.8246&2.199&0.9182
\\\hline$(sc)_{1^+}$&2.714&0.7469&2.518&0.8146&2.310&0.8776
\\\hline$(cc)_{1^+}$&3.909&0.6086&3.648&0.6360&3.381&0.6597
\\\hline$(ub)_{1^+}$&5.975&0.6235&5.823&0.7122&5.662&0.7935
\\\hline$(db)_{1^+}$&5.980&0.6426&5.828&0.7176&5.666&0.7935
\\\hline$(bs)_{1^+}$&6.158&0.6628&5.970&0.7165&5.771&0.8260
\\\hline$(bc)_{1^+}$&7.315&0.5350&7.048&0.5573&6.778&0.5812
\\\hline$(bb)_{1^+}$&10.66&0.4299&10.37&0.4373&10.08&0.4445
\\\hline
\end{tabular}
\begin{table}[h]
  \centering
  \caption{diquark mass(GeV) and mean square root radius (fm) with $\beta=0.75$}\label{tab7}
\end{table}
\end{center}
\vspace{0.3cm}
\begin{center}
\begin{tabular}{|c|c|c|c|c|c|c|}\hline
\multicolumn{1}{|c|}{$V_0(GeV)$}&\multicolumn{2}{|c|}{0.0}&\multicolumn{2}{|c|}{-0.3}
&\multicolumn{2}{|c|}{-0.6}\\\hline &mass&msr&mass&msr&mass&msr
\\\hline$(ud)_{0^+}$&-&-&-&-&-&-
\\\hline$(us)_{0^+}$&-&-&-&-&0.7448&9.8071
\\\hline$(ds)_{0^+}$&-&-&-&-&0.7903&9.8484
\\\hline$(uc)_{0^+}$&2.561&0.8500&2.454&0.8873&2.331&0.9410
\\\hline$(dc)_{0^+}$&2.568&0.8435&2.460&0.8814&2.335&0.9359
\\\hline$(sc)_{0^+}$&2.762&0.7208&2.617&0.7705&2.454&0.8285
\\\hline$(ub)_{0^+}$&6.070&0.7409&5.942&0.7918&5.803&0.8526
\\\hline$(db)_{0^+}$&6.077&0.7359&5.946&0.7869&5.806&0.8481
\\\hline$(bs)_{0^+}$&6.254&0.6418&6.087&0.6919&5.910&0.7490
\\\hline$(bc)_{0^+}$&7.417&0.4771&7.153&0.4883&6.906&0.5295
\\\hline$(uu)_{1^+}$&-&-&-&-&-&-
\\\hline$(ud)_{1^+}$&-&-&-&-&-&-
\\\hline$(dd)_{1^+}$&-&-&-&-&-&-
\\\hline$(us)_{1^+}$&-&-&-&-&0.7450&10.5665
\\\hline$(ds)_{1^+}$&-&-&-&-&0.7905&10.6108
\\\hline$(ss)_{1^+}$&0.9872&10.4853&1.385&10.8219&1.236&0.9149
\\\hline$(uc)_{1^+}$&2.606&0.5984&2.458&0.6957&2.300&0.7763
\\\hline$(dc)_{1^+}$&2.613&0.6016&2.464&0.6991&2.304&0.7776
\\\hline$(sc)_{1^+}$&2.807&0.6531&2.623&0.7178&2.427&0.7731
\\\hline$(cc)_{1^+}$&4.015&0.5604&3.760&0.5831&3.500&0.6037
\\\hline$(ub)_{1^+}$&6.048&0.5339&5.909&0.5870&5.760&0.6630
\\\hline$(db)_{1^+}$&6.056&0.5290&5.914&0.5896&5.765&0.6567
\\\hline$(bs)_{1^+}$&6.244&0.5782&6.067&0.6251&5.881&0.6807
\\\hline$(bc)_{1^+}$&7.414&0.4878&7.151&0.5127&6.886&0.5403
\\\hline$(bb)_{1^+}$&10.75&0.4052&10.46&0.4104&10.18&0.4166
\\\hline
\end{tabular}
\begin{table}[h]
  \centering
  \caption{diquark mass(GeV) and mean square root radius (fm) with $\beta=1.0$}\label{tab7}
\end{table}
\end{center}
\vspace{0.3cm}

In the approach, we take the parameters which are obtained by
fitting data for mesons \cite{zhang1,wang1}, then the only
parameters which can be adjusted are $\beta$ and $V_0$. There are
several remarks to make.

(1) The masses of $0^+$ and $1^+$ diquarks which are composed of two
heavy quarks or one-heavy-one-light quarks tend to be degenerate.
This is understandable in the heavy quark effective theory (HQET)
\cite{Isgur} and the results are  consistent with that given in refs
\cite{wilczek,konstantinos}.

(2) With the same $\beta$ value, the estimated masses of diquarks
increase as $V_0$ is larger.

(3) The masses of diquarks increase with increment of the $\beta$
value.

It is also observed from the numerical results, that for lighter
diquarks the axial diquark is 200 MeV heavier than the
corresponding scalar diqaurk  which is consistent with
\cite{konno} and  the lattice calculations \cite{konstantinos}. As
indicated, our approach might be suspicious for dealing with light
diquarks, however, consistency of the numerical results with that
obtained in other approaches indicates its limited plausibility.
We will discuss this issue in next section.

\section{Discussion and conclusion}

In this paper we systematically construct the wavefunctions of the
scalar and axial vector diquarks and then numerically evaluate
their spectra.

There are several free parameters which cannot be determined so
far, because the diquarks are not experimentally measurable. If
one tries to fix the parameters, he has to deal with the diquarks
which reside in baryons. However in that case, the diquarks are no
longer free. As the first step, in our work, we investigate the
free diquarks which cannot independently exist in real world, and
in our later works, we will take into account the effects due to
existence of the extra quark in baryon. We employ the
Bethe-Salpeter theory to study their spectra and some
characteristics. One of the free parameters is $\beta$ which
signifies the difference for the non-perturbative QCD confinement
effect between diquark and meson and from a naive consideration,
it should be within a range $0\leq\beta\leq 1.0$. Our numerical
results confirm that as $\beta>0.75$, there is no solution.
Another important parameter is $V_0$ which stands as the
zero-point energy (in the momentum space, it has a form of
$\propto V_0\delta^3({\bf q})$). In the case of mesons, we know
well that such zero-point energy must be introduced to meet the
data. In the case of diquark, it plays even more important role as
it severely determines the characteristics of the two-quark
system.

Moreover, as we apply the B-S approach to evaluate the diquarks of
two heavy quarks or one-heavy-one-light quarks, everything works
well, however, once we extend the same approach to the system of
two light quarks, the solutions do not seem to be reasonable, or
there are no solutions at all as the parameters take certain
values. It is not surprising, because as well known, it is hard to
use either the potential model or B-S equations to deal with pions
which are supposed to contain $q$ and $\bar q'$ with $q,q'$ being
$u,d$ quarks. It is believed that the quark masses in this case
are not simply a constant. The case about pions are carefully
analyzed by Dai et al. \cite{Dai} and the diquark system with two
light quarks are studied by Wang et al. \cite{Wang}.

Indeed the diquark is not color-singlet and not a real physical
state. It resides, generally, in baryons. Therefore, the estimated
mass and mean square-root radius of a free diquark may be
different from its practical value in baryons, because of the QCD
interaction of the extra quark with this subject (diquark), if it
indeed exists. That is the goal of our next work.

In the relativistic quark potential model,  Ebert et
al.\cite{ebert} obtained  the values of the masses and
$<r^2>^{1/2}$  as 3.226 GeV, 0.56 fm and 9.778 GeV, 0.37 fm for
$1^3S_1$ of the axial diquarks $cc$ and $bb$ respectively, with
inputs $m_c=1.55$ GeV and $m_b=4.88$ GeV. Comparing our results
with theirs, deducing the rest mass contributions ($2m_c$ and
$2m_b$) whereas we employ  larger input values for $m_c$ and $m_b$
in the numerical computations, our spectra are consistent with
theirs and the mean square-radius estimated in this work is about
1.2 times larger than that given in ref. \cite{ebert}. Considering
theoretical uncertainties, our results are qualitatively
consistent with the values of ref.\cite{ebert}.

For lighter scalar diquarks, many authors estimated their spectra
and obtained widely diverging results. For example, the authors of
\cite{ekelin,ekelin2,vogl} obtained $M_{(ud)_{0^+}} \leq 0.3GeV$,
whereas the authors of
\cite{hong,zlhdc,ram,hong3,semay,nagata,cristoforetti,santopinto,shuryak2,horikawa,schaefer}
got $M_{(ud)_{0^+}}=0.3-0.6GeV$, but $M_{(ud)_{0^+}}>0.6GeV$ was
suggested in Refs.
\cite{burden,glozman,marek,hess,tiburzi,ebert1,ebert2,cloet,nagata2,guo,karliner2}.
For $M_{(us)_{0^+}}$,  it is estimated as $0.63GeV$ \cite{hong},
$0.88GeV$ \cite{burden},   $0.948GeV$ \cite{ebert1} and 0.895GeV
\cite{ebert3}. For $M_{(ud)_{1^+}}$, it is estimated as $0.614GeV$
\cite{ram} , $0.806\sim 0.95GeV$
\cite{konstantinos,glozman,ebert1,ebert3,burden}, $1.05 GeV\sim
1.27 GeV$ \cite{nagata,nagata2,cloet}, $>1.27GeV$ \cite{cahill}.
For $M_{(us)_{1^+}}$, $1.069GeV$ \cite{ebert1} and $1.5GeV$
\cite{ebert3}. For $M_{(ss)_{1^+}}$, it is estimated as
$1.203GeV$\cite{ebert1} and $1.215$ GeV \cite{ebert3}. The mass of
$(cu)_{0^+}$ is given as $1.933GeV$ \cite{maiani}. The various
numbers indicate large theoretical uncertainties. To fix them, one
needs to apply the results to baryons which is compose of diquarks
and are observable physical states.

As suggested, there could be the 1/2-rule that all the
corresponding values and probably $V_0$ employed for the diquarks,
i.e. $\beta$ and $V_0$ should be 1/2 of the values for mesons.
This rule follows the assumptions that the governing interaction
is QCD, thus both short-distance (the Coulomb) and the
long-distance (the confinement) are proportional to the
expectation value of the Casimir operator $<\lambda^a\lambda^a>$
where $\lambda^a$ is the SU(3) Gell-Mann matrix. If so, the set of
results with $\beta=0.5$ should be more reasonable. However, the
situation may be more complicated. Thus in this work, we keep it
as an adjustable free parameter. As we indicated above, the
diquark is not a physical state, so that to fix the parameter we
need to put them into the baryons, or may be in the future, as
hoped,  in the high-energy heavy ion collisions, free diquark
might be directly produced, and then we can have more information
about its structure.

Indeed, until we know how  the diquark interacts as a whole object
with gluons, we cannot more reliably estimate the baryon case and
evaluate the production rate and decay width of the baryon in the
quark-diquark picture. Therefore, in our coming work, we are going
to derive all the form factors at the effective interaction
vertices as the diquarks are treated as
a whole object.\\

\noindent Acknowledgement: This work is partly supported by the
National Natural Science Foundation of China (NNSFC) and The
Special Fund for the Ph.D programs of Chinese Universities. We
highly benefit from discussions with Prof. C.-H. Chang who kindly
introduces their new methods for numerically solving the B-S
equation to us and indicates some important physics problems which
we did not notice.

\end{document}